\documentstyle[12pt,aaspp]{article}
\begin{document}

\title{COLOR-MAGNITUDE DIAGRAM DISTRIBUTION OF THE BULGE RED CLUMP STARS --
EVIDENCE FOR THE GALACTIC BAR}
\author{K. Z. Stanek\altaffilmark{1}}
\affil{Princeton University Observatory, Princeton, NJ 08544--1001}
\affil{\tt e-mail I: stanek@astro.princeton.edu}
\altaffiltext{1}{On leave from N. Copernicus Astronomical Center,
Bartycka 18, Warszawa 00-716, Poland}
\author{M. Mateo}
\affil{Department of Astronomy, University of Michigan, 821 Dennison
Bldg., Ann Arbor, MI~48109--1090}
\author{and}
\author{A. Udalski, M. Szyma\'nski, J. Ka\l u\.zny, M. Kubiak}
\affil{Warsaw University Observatory, Al. Ujazdowskie 4, 00--478 Warszawa,
Poland}

\begin{abstract}
The color-magnitude diagrams of $ \sim 5 \times 10^5 $ stars
obtained for 13 fields towards the Galactic bulge with the OGLE
project reveal a well-defined population of bulge red clump stars.
We find that the distributions of the extinction-adjusted apparent
magnitudes of the red clump stars in fields lying at $l=\pm5\deg$
in galactic longitude differ by $0.37\pm0.025\;mag$.
Assuming that the intrinsic luminosity distribution of the red clump stars
is the same  on both sides of the Galactic center, this implies that
the distances to the red clump stars in the two fields differ by
a factor of $1.185\pm0.015$.
A plausible explanation of the observed difference in the luminosity
distribution is that the Galactic bulge is a triaxial structure, or bar,
which is inclined to the line of sight by no more than $45\deg$,
with the part of the bar at the positive galactic longitude being
closer to us. This agrees rather well with other studies indicating
the presence of the bar in the center of the Galaxy.
Color-magnitude diagram data are accessible over the
computer network with anonymous {\tt ftp}.

\end{abstract}


\section{INTRODUCTION}

The Optical Gravitational Lensing Experiment (OGLE, Udalski et al.~1992,
1993b, 1994) is an extensive photometric search for the rare cases of
gravitational microlensing of Galactic bulge stars
by foreground stars, brown dwarfs and planets.
It provides a huge data base (Szyma\'nski \& Udalski 1993),
from which color-magnitude diagrams have been compiled (Udalski et al.~1993a).
These color-magnitude diagrams reveal an expected population of bulge stars,
with its turn-off
point, red giant branch and red clump, and also an unexpected
concentration of stars in the blue part of the color-magnitude diagrams,
which was recently explored in detail  by Paczy\'nski et al.~(1994).
In this paper we use the  well-defined population of red clump
stars to investigate the presence of a triaxial structure, or bar,
in the bulge of the Galaxy.

A number of recent studies show unambiguously the presence of such a
structure in the center of the Galaxy. Blitz \& Spergel (1991) analyzed
$2.4\;\mu m$ observations of the Galactic center of
Matsumoto et al.~(1982) and showed convincingly that the observed
asymmetry in the galactic longitude distribution of surface brightness
is naturally explained by the bar with the near side in the first Galactic
quadrant. They also argue that there is a small tilt of the bar with
respect to the Galactic plane, consistent with the tilt proposed by Liszt
\& Burton (1980) from $21\;cm$ emission kinematics. Binney et al.~(1991)
have constructed a dynamical model for the HI, CO and CS emission
in the inner Galaxy, and their resulting bar has the same orientation
as that suggested by Blitz \& Spergel~(1991) in the sense that the  closer
part of the bar is at positive galactic longitudes. COBE multiwavelenght
observations of the Galactic center (Weiland et al.~1994) confirmed
the existence of the longitudinal asymmetry discussed by Blitz
\& Spergel~(1991), but Weiland et al.~argue that the signature of the
tilt disappears when the Galactic center emission is corrected for absorption.

There is also evidence for the triaxial structure in the center of the Galaxy
from star counts. Nakada et al.~(1991) analyzed the distribution of
IRAS Galactic bulge stars and found asymmetry in the same sense as
Blitz \& Spergel~(1991) and Binney et al.(1991). Whitelock \&
Catchpole~(1992) analyzed the number distribution of Miras in the bulge
as a function of distance modulus and found that the half of the bulge
which is at positive galactic longitude is closer to us than the other
half. The observed stellar distribution could be modelled with
a bar inclined at roughly $45\deg$ to the line of sight.
Weinberg~(1992) used AGB stars as star tracers and mapped the Galaxy
inside the solar circle. He found evidence for a large stellar
bar with semimajor axis of $\approx\;5\;kpc$ and inclination
placing the nearer side of the bar at positive galactic longitudes.
Spergel~(1992) gives a detailed account of the evidence for the bulge
being barred.

\section{THE DATA}

Udalski et al.~(1993a) present color-magnitude diagrams (CMDs) of
14 fields in the direction of the Galactic bulge,
which cover nearly one square degree and contain about $5 \times 10^5$
stars. All observations
were made using the 1 meter Swope telescope at the Las  Campanas Observatory,
operated by the Carnegie Institution of Washington, and a $ 2048 \times 2048 $
pixel Ford/Loral CCD detector with the pixel size 0.44 arcsec covering
$15' \times 15'$ field of view.   In this paper we discuss nine
fields in Baade's Window (BW) and four fields on both
sides of the Galactic center (MM5, MM7). Table~1 gives the coordinates
of four MM fields analyzed in this paper and Figure~1 shows schematically
the positions of all 13 fields in galactic coordinates.
As an example, the CMD  for one of the positive galactic latitude
fields (MM7-A) is shown in Figure 2, together with two straight lines
corresponding to values of 11.5 and 13.0 for the extinction-insensitive
parameter $V_{_{V-I}}$ (cf.~eq.~\ref{eq:free} in this paper).
Most of the diagram is dominated by bulge stars, with red clump
stars lying approximately between the two lines shown.
The part of the diagram dominated by the disk stars was recently analyzed by
Paczy\'nski et al.~(1994). In this paper we use well-defined population of
bulge red clump stars to investigate the presence of the
bar in the center of the Galaxy. Red clump stars are the equivalent of
the horizontal branch stars for metal rich population, i.e.
relatively low mass stars burning helium in their cores.
{}From observations and also from stellar evolution theory (Castellani, Chieffi
\& Straniero 1992) we expect the bulge red clump
stars to be relatively bright and have a narrow luminosity distribution
with weak dependence on the  metallicity. Therefore, red clump stars
form a suitable population with which to investigate the structure of
high-metallicity systems, like the  Galactic bulge. As we observe several
thousand red clump stars in each field, we expect they can
be used as a powerful tool in investigating the structure of the bulge.
The part of the CMD dominated
by bulge red clump stars is shown for MM5 and MM7 fields
in Figure~3, with the same two straight lines as in Figure~2.
It is clearly visible that the red clump stars from the MM7 fields
group close to the $V_{_{V-I}}=11.5$ line, while red clump stars
from the MM5 fields have, on average, larger values of this parameter.

To analyze the distribution of bulge red clump stars in a more quantitative
manner, we define the extinction-insensitive $V_{_{V-I}}$ parameter
\begin{equation}
  V_{_{V-I}} \equiv V - 2.6 ~ (V-I),
\label{eq:free}
\end{equation}
where we use reddening law $ E_{_{V-I}} = A_{_V}/2.6 $,
following Dean, Warren, \& Cousins (1978) and Walker (1985).
The parameter $ V_{_{V-I}} $ has been defined so that if
$A_{_V}/E_{_{V-I}}$ is independent of location then
for any particular star its value is not affected by the unknown
extinction.  It was found by Paczy\'nski et al.~(1994) that the
distribution of the peaks of the $ V_{_{V-I}} $ parameter has deviation from
the mean as small as $0.03 $ in all nine BW fields, indicating that it
is insensitive to the interstellar extinction, as designed.
Then for all 13 fields we consider only the region of the CMD clearly
dominated by the bulge red clump stars:
\begin{equation}
15 < V < 19.0 ~~;~~~ 1.5  < V-I < 2.4 .
\label{eq:select}
\end{equation}
All stars observed in nine fields in Baade's Window,
two fields in the MM5 window, and two fields in the MM7 window that satisfied
the inequalities (\ref{eq:select}) were put into three separate
data sets and counted in bins of $ \Delta V_{_{V-I}} = 0.05 $.
The result appears in Figure~4, where we see the number of stars
as a function of $V_{_{V-I}}$ for MM7, BW and MM5.
The histogram for the red clump stars in nine BW fields was
scaled by dividing the total number of stars by 6 as to obtain
approximately equal peak number as in two MM5 fields and two MM7 fields.
For the purpose of presentation  the distributions
were boxcar-smoothed with a filter width of three bins.
Also shown is the expected contamination of the selected CMD region
by disk stars in BW field (normalized to the same area as contained
by MM fields), calculated using standard Bahcall \& Soneira~(1980)
model of the galaxy. Clearly this contamination can be safely neglected.

Distributions shown in Figure~4 are similar in shape,
with red clump stars forming a pronounced peak in observed distributions.
There is however a clear shift between the distributions, with
MM7 red clump stars having on average smallest values of $V_{_{V-I}}$
parameter and MM5 red clump stars having largest values of $V_{_{V-I}}$
parameter, with BW stars in between. To quantify this shift
in more detail, we applied the iterative bootstrap technique
(for the  discussion of the bootstrap method see Press et al.~1992).
First, we estimated the shift between the distributions by eye and we
selected for all three fields the same region of distributions, which
for BW7 field corresponded to $11.0 < V_{_{V-I}} < 13.5$.
The region of comparison was asymmetric with respect to the peak value of
$V_{_{V-I}}$ distribution so as to avoid those values of $V_{_{V-I}}$ which are
contaminated by bulge red giants and also, to a lesser extent, by disk stars.
Then for every field using bootstrap selected samples
we estimated the mode of the distribution (Lupton~1993)
and obtained the shift between the distributions, which we then applied
to correct the comparison region of $V_{_{V-I}}$ distributions.
The resulting plot of $\Delta V_{_{V-I}}$ for 10,000 Monte Carlo bootstrap
selected data sets is shown in Figure~5. We find that the distribution of
shift $\Delta V_{_{V-I}}$ is very well fitted by gaussian, with parameters
$\Delta V_{_{V-I}}({\rm MM5-BW})=0.15\pm0.02$
for BW and $\Delta V_{_{V-I}}({\rm MM5-MM7})=0.37\pm0.025$ for MM7.

There is an additional quantity one can obtain from our data.
This is the density  of red clump stars for different fields.
We find that there were $\sim45,740$ red clump stars in nine BW fields,
$\sim7,280$ in two MM7 fields and $\sim7,540$ stars in two MM5 fields,
satisfying inequalities (\ref{eq:select}) and falling, after shift, within
the comparison region mentioned above. This corresponds to a number
of red clump stars per field ($15'\times15'$) of 5,080 for BW, 3,640 for
MM7, and 3,770 for MM5. In the following section we will discuss
the implications of our observations for the structure of the
Galactic bulge.

\section{DISCUSSION}

In previous section we have shown that the distributions of
bulge red clump stars, located on both sides of the Galactic center,
as a function of extinction-adjusted apparent
magnitude  are very similar in shape but differ by substantial
shift which was found to be $\Delta V_{_{V-I}} ({\rm MM5-MM7})=0.37\pm0.025$.
One possible explanation for this shift is a difference in the reddening
law (see discussion following Eq.1) for different fields in the Galactic
bulge. If the ratio of $A_{_V}/E_{_{V-I}}$ for the BW field is 2.6,
as shown by Paczy\'nski et al.~(1994), then one needs this ratio to be
about $\sim2.9$ for the MM5 field and $\sim2.3$ for the MM7 field,
with exact values depending on the $E_{_{V-I}}$ value for BW field.
We find such a large differences rather unlikely, especially
considering the small distances between the fields,
but we expect to  address this question in more detail in the future.
If we attribute the observed shift as being due to the difference
in distance to the bulge red clump stars in MM5 and MM7 then we can obtain
the ratio of distances to both fields $d_1/d_2=1.185\pm0.015$.
If we then assume that the observed peaks in the $V_{_{V-I}}$ distributions
correspond to the stars lying along major axis of the bar, we can obtain
the angle of inclination of the bar to the line of sight
$\theta\approx45\deg$.
To check how this angle corresponds to the real inclination of the bar to the
line of sight, we modeled the bar with Blitz \& Spergel~(1991)
Eq.1, taking $x_s=1\;kpc, z_s=0.6\;kpc$ and changing $y_s$
from $0.05$ to $1.0\;kpc$. We then calculated how the observed inclination
changes with increasing thickness of the bar. Figure~6 shows the
result for three values of intrinsic inclination $15, 30$ and $45\deg$
($\theta=90\deg$ corresponds to the major axis of
the bar being perpendicular to the line of sight).
For a bar very thin along the line of sight the inclination angle as
measured corresponds directly to the true inclination angle.  However,
if the bar is thick then the true inclination angle is smaller than
the angle measured on the basis of the mean distance to stars in the
fields MM5 and MM7.
We suspect that the tendency of the observed angle to be always
greater than the intrinsic value is the generic feature of all
realistic models of the bar. So, at present we can only safely
state that there is substantial asymmetry in the distance to the red clump
stars on both sides of the Galactic center, strongly indicating the presence
of the bar in the Galactic center.

Star counts can also provide very useful, direct information about
space distribution of star density along the line of sight. In Figure~4
notice that red clump stars $V_{_{V-I}}$ distribution is
relatively narrow ($FWHM\approx1.0\;mag$), which  gives us an upper limit
for the spatial extent of bar red clump stars along the line of sight
of about $4\;kpc$. To obtain more stringent
limitations one needs some additional knowledge about intrinsic luminosity
distribution of red clump stars.

We also see that for BW fields
there is about 40\% more red clump stars than in the
MM fields. At the distance of
Galactic center, which we assume to be $8\;kpc$, $5\deg$ corresponds
to about $700\;pc$ in the plane of the sky, or to about $1\;kpc$
if we apply $\sim45\deg$ inclination discussed above. This tells us that the
bar major axis scale length is comparable to $1\;kpc$, but to obtain a
better estimate for this value we need additional fields with larger
values of $|\,l\,|$. The Galactic bar observed by COBE can still be  seen
at $l=\pm15\deg$ (Weiland et al.~1994).

The presence of the bar in the Galaxy seems to be firmly established
by various authors and methods (Blitz \& Spergel~1991; Binney et al.~1991;
Nakada et al.~1991; Weinberg 1992; Whitelock \& Catchpole~1992;
Weiland et al.~1994), but there are still
considerable differences as to details of the bar structure
or angle of inclination to the line of sight. We have shown that
the red clump stars can be very useful for investigating
Galactic bar, being both numerous and relatively bright.
We expect to address this problem in the future with data covering
much  wider range of galactic coordinates.

We also note  here the possibility that the Galactic bar
may be associated with the  deficiency of Galactic disk stars
beyond $\sim 3\;kpc$ from the Sun towards the Galactic bulge
(Paczy\'nski et al.~1994), as compared with standard models of the Galaxy.
The disks of barred galaxies often show a decrease in brightness interior to
the end of the bar (Kormendy~1994). This effect was recently observed in the
near-infrared by Spillar et al.~(1992) in the galaxy NGC 5195.
A similar decrease may exist in our Galaxy, although such a deficiency
also occurs in inner disks of non-barred galaxies
(Freeman~1970; Kormendy~1977).

This paper and distribution of stars in the color-magnitude
diagram as observed by OGLE in BW, MM5 and MM7 fields is available over the
computer network using  anonymous {\tt ftp} on {\tt astro.princeton.edu}.
Login as {\tt ftp}, use your name as a password.  Change directory to
{\tt stanek/bar}. The file {\tt read.me} contains a list of the necessary
files and instructions how to retrieve the data.

\acknowledgments{We would like to thank B.~Paczy\'nski,
the PI of the OGLE project, for encouragement, many stimulating
discussions and comments. We acknowledge comments from J.~E.~Rhoads,
D.~N.~Spergel and N.~D.~Tyson, who read an earlier version of this paper,
and also comments from the anonymous referee, which allowed us to
improve the final version of this paper.
We also acknowledge discussions with R.~Lupton
and J.~E.~Gunn. This project was supported with the NSF grants AST 9216494
and AST 9216830 and Polish KBN grants No 2-1173-9101 and BST438A/93.}

\begin{figure}
\begin{center}
{\bf FIGURE CAPTIONS}
\end{center}

%Fig. 1
\caption{Positions in the Galactic coordinates of 13 fields analyzed
in this paper, for which the  $V-I$ color-magnitude diagrams
were obtained by the OGLE experiment (Udalski et al.~1993a, see also Table~1).}

%Fig. 2
\caption{The $V-I$ color-magnitude diagram for stars in the
MM7-A field of the OGLE experiment (Udalski et al.~1993a).
The two straight lines correspond to the value of extinction-free
parameter (Eq.~1) $V_{_{V-I}}$ equal to 11.5 and 13.0.}

%Fig. 3
\caption{Region of the $V-I$ color-magnitude diagrams
dominated by bulge red clump stars for  four
MM fields of the OGLE experiment (Udalski et al.~1993a).
Stars were selected to satisfy the inequalities given by Eq.~2.
As in Figure 2, the two straight lines correspond to the value of
extinction-insensitive parameter $V_{_{V-I}}$ equal to 11.5 and 13.0.}

%Fig. 4
\caption{Histograms of the $V_{_{V-I}}$ distribution for red clump stars
from MM5 (continuous line), BW (dotted line) and MM7 (short-dashed line).
The histogram for the red clump stars in BW was
normalized by dividing the total number of stars by 6 so as to obtain
approximately the same peak number as in MM5 and MM7.
For the purpose of presentation  the distributions
were boxcar-smoothed with a filter width of three bins.
Also shown is estimated contamination from disk stars at the red clump,
based on the Bahcall \& Soneira~(1980) model
of the Galaxy (long-dashed line).}

%Fig. 5
\caption{Plot of the $\Delta V_{_{V-I}}$ shift distribution
for MM7 field (continuous line) and BW field
(dashed line) versus MM5 field. For details see text.}

%Fig. 6
\caption{Observed angle of inclination $\theta_{obs}$ (continuous line)
as a function of bar axis ratio $y_s/x_s$.
The bar was modeled using Blitz \& Spergel~(1991)
Eq.~1, with fixed $x_s=1.0\;kpc, z_s=0.6\;kpc$ and changing value
of $y_s$. Three values of the intrinsic inclination $\theta_0=15, 30, 45\deg$,
shown with dashed lines, were investigated.}

\end{figure}

\newpage
\titlepage

\begin{planotable}{lrrr}
\tablewidth{20pc}
\tablecaption{Parameters for the four MM fields}
\tablehead{
\colhead{field}	& \colhead{$l$}	& \colhead{$b$} &
\colhead{CMD}  \vspace{0.1cm} \\
\colhead{} 	& \colhead{[$\,\deg\,$]} & \colhead{[$\,\deg\,$]}
& \colhead{stars} }

\startdata
MM5-A	& $-$4.77 & $-$3.36 & 28077 \nl
MM5-B	& $-$4.94 & $-$3.46 & 28906 \nl
MM7-A	& 5.43 	  & $-$3.34 & 30298 \nl
MM7-B	& 5.53    & $-$3.52 & 45480 \nl
\end{planotable}

\end{document}